\documentstyle[graphicx,float]{article}

\begin{document}
\title{Revisiting entanglement-induced exclusion}
\date{}
\author{Pedro Sancho \\ Centro de L\'aseres Pulsados CLPU \\ Parque Cient\'{\i}fico, 37085 Villamayor, Salamanca, Spain}
\maketitle
\begin{abstract}
We revisit the problem of absorption by identical atoms with entanglement and symmetrization acting at once. We introduce the recoil of the atoms in the calculations and compare the results with those of distinguishable systems instead of mixtures. These absorption rates have been proposed to test entanglement-induced exclusion. The manifestation of exclusion in this framework is a sharp increase of the absorption rates in its proximity. The control of the superposition coefficients used in this representation is simpler than that of the overlapping degree considered in the initial proposal.  Another result of this alternative approach is the inhibition of the entanglement effects for some values of the parameters.  
\end{abstract}

\section{Introduction}

Entanglement and symmetrization are two forms of non-separability present in systems of identical particles. The interplay between them is a very subtle subject. For instance, there is not yet consensus about the correct definition of entanglement in systems of identical particles \cite{tri}.  A related topic, the possibility of converting symmetrization into useful entanglement, has been only settled in the last years, when some protocols showing how to implement this task have been proposed \cite{san,ple,lof}. 

More recently, another two aspects of the above interplay has been discussed in the literature. The first one is the analysis of the joint effects of entanglement and symmetrization in atomic systems when both are simultaneously relevant \cite{yap}.  In that reference the absorption rates of identical atoms in a symmetrized multi-atom superposition were compared with those of a mixture for several values of the spatial overlap between the atoms. When both types of non-separability are present at once we observe large deviations with respect to the behaviour of only entangled or symmetrized states and of mixtures. 

The second aspect is the possibility of entanglement-induced exclusion. Standard exclusion, based on Pauli's principle \cite{Pau}, refers to the impossibility of preparing identical fermions in the same state. In the case of entangled particles it is impossible to define pure individual states for each particle and, consequently, we cannot apply the above principle. We must use different criteria, not based on the one-particle states of the multi-fermion system, to determine potential excluded states \cite{yar}. Interestingly, in \cite{yap} a realistic example of non-standard excluded state was found. 

We revisit the above scheme by considering some modifications that lead to a more complete mathematical description, a more clear visualization of the differences between bosons and fermions, and a simpler identification of the exclusion behaviour. In the original calculation of the absorption rates, the recoil effect in the atoms was not taken into account. This effect can be important because it changes the spatial overlap between the atoms after the absorption. A correct evaluation of the overlapping degree is necessary to correctly determine the strength of the exchange effects in the system. A second modification is related to the control parameters in the problem. In \cite{yap} the absorption rates were presented as a function of the overlap between different spatial states of the atoms. Here, instead, the control parameters will be the coefficients of the multi-fermion superposition (determining the entanglement degree). This approach has the fundamental advantage of being simpler to implement. These coefficients can be tuned in an easier way than the atomic overlaps. The third difference with the original proposal is that there the comparison of the absorption rates of entangled identical atoms was done with those of mixtures of the same atoms. Here, at variance, the comparison is  with pairs of distinguishable atoms having a similar degree of entanglement. In this complementary view we can show more clearly the identity effects when the system is already entangled. 

As a by-product of the above analysis we derive another interesting result, the inhibition of the dependence of the absorption rates on the coefficients of the multi-fermion superposition for some parameters of the problem. This cancellation of the entanglement effects is a novel manifestation of the strong interplay between both forms of non-separability.    

\section{The physical system}

We consider the same arrangement described in \cite{yap}. It is based on the absorption properties of two identical atoms generated in the photodissociation of molecules by light in a superposition state. Molecular photodissociation has been previously used  to experimentally study the influence of entanglement on the process of spontaneous emission  \cite{jap,bel}. As the identity of the atoms must also be taken into account in order to explain the results of the experiments \cite{ypr}, this process is well suited to carry out tests where entanglement and identity are relevant at once.   

A molecule composed of two atoms, distinguishable or identical,  is photo-dissociated into two atoms in a superposition state. To prepare the atoms this way the molecule must interact with light in a superposition state, giving rise each term in the superposition to a different dissociation process. As explained in \cite{yap} this can be achieved dividing the light beam with a beam splitter and then having different incident directions on the molecule. The states after the dissociation for distinguishable and identical atoms are respectively
\begin{equation}
|\Psi>=N_D(a|\psi >_A|\phi >_B +b|\varphi>_A|\chi >_B)|g>_A|g>_B
\label{eq:uno}
\end{equation}    
and
\begin{eqnarray}
|\Phi>=N_I[(a(|\psi >_1|\phi >_2 \pm  |\phi >_1|\psi >_2   )+ \nonumber \\
b(|\varphi>_1|\chi >_2 \pm |\chi>_1|\varphi >_2)]|g>_1|g>_2
\label{eq:dos}
\end{eqnarray}    
where $a$ and $b$ are the coefficients of the superposition, obeying the relation $|a|^2+|b|^2=1$. In the case of distinguishable particles we use the subscripts $A$ and $B$ to denote the atoms, whereas for identical ones we use $1$ and $2$. This way we emphasize that the labels have very different meanings in both cases. In the first one, they can be associated with distinguishable physical properties of the atoms, whereas in the second one they are only mathematical labels without relation to properties of each atom.  The states $\psi$, $\phi$, $\varphi$ and $\chi$ represent the spatial centre of mass (CM) states of the atoms.  Because they are generated by dissociation there is strong spatial overlapping between them. In the case of distinguishable particles we have $<\psi|\varphi>_A \neq 0$ and $<\phi |\chi>_B \neq 0$, that is, we have two non-orthogonal states for each particle. In contrast, for identical systems the scalar products between the four states make sense because the particle labelled $i$ ($i=1,2$) can be in the four states: $<\psi |\varphi >_i \neq 0$,  $<\psi |\phi >_i \neq 0$, ... Note that the presence of overlapping between the identical atoms is a necessary condition for the  existence of exchange effects. In the double sign for identical atoms the upper one corresponds to bosons and the lower one to fermions. On the other hand, the atoms are in their ground internal state, which is denoted by $g$. Finally, the normalization coefficients are
\begin{equation}
N_D=(1+2Re(a^*b<\psi|\varphi><\phi|\chi>))^{-1/2}
\end{equation}   
and 
\begin{eqnarray}
N_I=(2 \pm 2|a|^2|<\psi |\phi>|^2 \pm 2|b|^2|<\varphi |\chi>|^2+ \nonumber \\
4Re(a^*b <\psi |\varphi><\phi |\chi>) \pm 4Re(a^*b <\psi |\chi><\phi |\varphi> )  )^{-1/2}
\end{eqnarray}
An important characteristic of our system is that in the distinguishable case the states of each particle are non-orthogonal. This contrast with most studies on entanglement, which consider orthogonal states. We can study the influence of non-orthogonality on the modifications that entanglement  induces on the physical properties of the system. 

The state $\Phi$ is the antisymmetrization of $\Psi$ for a system of identical particles.  It contains simultaneously entanglement and identity effects. The first ones, associated with the two-atom superposition, can be seen as inherited from the state previous to the antisymmetrization. This intuitive view of entanglement in systems of identical particles is very in line with the approach advocated in \cite{ghi}, where the symmetrization process does not contribute to the entanglement degree. Anyway, the results of this paper are independent of this intuitive picture.  On the other hand, the identity effects also appear in Eq. (\ref{eq:dos}) as additional terms in the two-atom superposition. 

\section{Absorption rates}

After the preparation of the pair of atoms in a superposition state they interact with a light beam. In the case of the distinguishable particles the light beam is a broad band one containing the two absorption frequencies of the two atoms. The intensity of the beam is very low, making the possibility of two-photon absorption by each atom negligible. Similarly, the process of simultaneous double absorption, one by each atom can also be neglected. We only need to study the probability of single absorption.

As signalled in the Introduction, we want to analyse the dependence of the absorption rate on the degree of superposition. This view is complementary to that in \cite{yap}, where the focus was on the dependence on the overlapping between the atoms.  In addition we want to compare the rates of identical and distinguishable particles. In \cite{yap} the comparison was with a mixed state of the same identical atoms. The advantage of confronting the rates of identical and distinguishable atoms is specially clear when we consider for the second case an atom and one of its isotopes. Then we expect the dynamical interactions between both atoms to be very similar in both cases and, consequently, the differences in the entanglement present in both systems should be small. This can make the effects of identity more transparent. 

First, we evaluate the single absorption rate for distinguishable atoms.  The derivation is similar to that in \cite{yap} for identical atoms. We shall only introduce a change with respect to that calculation. In \cite{yap} we did not take into account the recoil of the  atoms, which we shall include now in the calculations.
   
The photon can be absorbed by atom $A$ or by atom $B$. The absorption by atoms in a superposition state leads to a new superposition, now also containing excited states. As it is well-known , the absorption process does not break a pre-existent superposition. This property strongly contrast with the behaviour of spontaneous emission, which can disentangle a multi-particle system \cite{ebe}. It is convenient to express the resulting superposition in terms of the not normalized excited states
\begin{equation}
|A^*>=(a|\psi ^*>_A|\phi >_B +b|\varphi^*>_A|\chi >_B)|e>_A|g>_B
\end{equation}  
and
\begin{equation}
|B^*>=(a|\psi>_A|\phi^* >_B +b|\varphi>_A|\chi^* >_B)|g>_A|e>_B
\end{equation} 
where $\psi ^*$ denotes the CM wave function after the recoil associated with the absorption, and $e$ is the excited state of the atom.  The states $A^*$ and $B^*$ represent the absorption of the photon by atom $A$ or $B$. Then the state after the absorption is
\begin{equation}
|\Psi ^*>=N_*(|A^*>+|B^*>)
\end{equation} 
The normalization coefficient is
\begin{equation}
N_*=(2+2Re(a^*b<\psi^*|\varphi^*><\phi|\chi>)+ 
2Re(a^*b<\psi|\varphi><\phi^*|\chi^*))^{-1/2}
\end{equation}
The absorption matrix element in the distinguishable case becomes 
\begin{equation}
{\cal M}_D = <0_{EM}|<\Psi^*|\hat{U}|\Psi>|1_{EM}>
\end{equation}
with $\hat{U}$ the evolution operator of the system and $0_{EM}$ and $1_{EM}$ the states of the electromagnetic field with $0$ and $1$ photons.

As we are considering low intensity light beams we can resort to perturbation theory in the usual electric-dipole approximation \cite{lou}. Because there is no interaction between the atoms their dynamics are independent and their interactions with the electromagnetic field are also independent. Then the Hamiltonian of the interaction between the field and the two-atom system (including the internal and CM variables) can be written as $\hat{H}=\hat{H}_A \otimes \hat{I}_B + \hat{I}_A \otimes \hat{H}_B$ with $\hat{H}_i$ ($i=A,B$) the one-atom interaction Hamiltonian and $\hat{I}_i$ ($i=A,B$) the identity operator. The explicit form of $\hat{H}_i$ will be discussed later. The evolution operator at any time $t$  is $\hat{U}(t)=\exp (-it\hat{H}/\hbar)$.  We can expand this equation as $\hat{U}(t) = \hat{I}-it\hat{H}/\hbar+\cdots$.  At first order of perturbation theory we have $\hat{U}(t) \approx \hat{U}_{first}(t) =\hat{I}- (it/\hbar)  (\hat{H}_A \otimes \hat{I}_B + \hat{I}_A \otimes \hat{H}_B)$. We could reach the same result noting that the no-interaction form of the Hamiltonian implies that the evolution operator factorizes, $\hat{U}=\hat{U}_A \otimes \hat{U}_B$, and then expanding the one-atom evolution operators $\hat{U}_A(t) = \hat{I}_A-it\hat{H}_A/\hbar + \cdots$. We assume $t$ to be fixed for all the calculations, a necessary condition to correctly define the absorption rates, which have the form  $|{\cal M}_D|^2/t$ \cite{lou}. 

Using the above expressions we can evaluate ${\cal M}_D$.   In ${\cal M}_D$ there are eight different terms. Let us calculate the first of them:
\begin{eqnarray}
N_*a^* <0_{EM}| _A<\psi^*|_A<e|_B<\phi|_B<g|\hat{U}aN_D|\psi>_A|g>_A|\phi>_B|g>_B|1_{EM}> \nonumber \\
\approx -it\hbar ^{-1}N_*N_D |a|^2 <0_{EM}|_A<\psi ^*|_A<e|\hat{H}_A|\psi>_A|g>_A|1_{EM}> 
\label{eq:die}
\end{eqnarray}   
In the deduction of this equation we have introduced the first order approximation of the evolution operator. Its first term, $\hat{H}_A \otimes \hat{I}_B$, leads to the transition matrix element of atom $A$ and to a trivial evolution for $B$, $<g|<\phi |\hat{I}_B|\phi>|g>=1$. On the other hand, the term $\hat{I}_A \otimes \hat{H}_B$ is null, $<e|<\psi ^* |\hat{I}_A|\psi>|g>=0$. 

Next, we must evaluate the matrix element in the r. h. s. of Eq. (\ref{eq:die}). As signalled before, the interaction Hamiltonian $\hat{H}_A$ is the electric-dipole one: the light-matter interaction is represented for each atom by $\hat{H}_i^{ED}$ ($i=A,B$). Using the transition operators $|e><g|$ and $|g><e|$ the Hamiltonian in the rotating-wave approximation can be expressed as \cite{lou}:
\begin{equation} 
\hat{H}^{ED}=G(|e><g|  \hat{a} \,  e^{i{\bf k}.{\bf R}}- |g><e|\hat{a}^+e^{-i{\bf k}.{\bf R}})
\end{equation}  
In this equation we have $G=G_0 {\bf e} . {\bf D}_{eg}$ with ${\bf e}$ the unit polarization vector. This vector is contained in the electromagnetic field operator, which is proportional to ${\bf e}(\hat{a} e^{i{\bf k}.{\bf R}}- \hat{a}^+e^{-i{\bf k}.{\bf R}})$, with $\hat{a}$ ($\hat{a}^+$) the annihilation (creation) operator, ${\bf R}$ the CM position, and ${\bf k}$ the wave-vector of the mode of the light field under consideration  (we have assumed that each atom only interacts with a single mode, being not necessary to include the summation associated with all the modes). On the other hand, ${\bf D}_{eg}$ is the expectation value of the total electric-dipole moment operator of the atom between the states of the transition,  ${\bf D}_{eg} =<e|\hat{\bf D}|g>$. Finally, $G_0$ contains all the constant coefficients appearing in the interaction Hamiltonian. 

The above form of the interaction Hamiltonian does not explicitly take into account the CM variables. They can be easily included in the picture by considering the action of the exponential term on the CM wave functions. In the case of an atom with CM momentum ${\bf p}$ we have
\begin{equation}
e^{i{\bf k}.{\bf R}}\frac{1}{(2\pi \hbar)^{3/2}} e^{i{\bf p}.{\bf R}/\hbar}=\frac{1}{(2\pi \hbar)^{3/2}}e^{i({\bf p}+\hbar {\bf k}).{\bf R}/\hbar} 
\end{equation}
that is equivalent in terms of states to $\exp(i{\bf k}.{\bf R})|{\bf p}> =  |{\bf p}+\hbar {\bf k}>$  \cite{cas}. In the case of the other exponential of the Hamiltonian we obtain $\exp(-i{\bf k}.{\bf R})|{\bf p}> =  |{\bf p}-\hbar {\bf k}>$.  It is simple to see that a similar result holds when the atom is in a superposition of momentum states. 

The action of the exponential on the CM states can also be expressed in terms of transition operators as $\exp(i{\bf k}.{\bf R})|{\bf p}> =  (|{\bf p}+\hbar {\bf k}><{\bf p}|)|{\bf p}>$. Then returning to our initial notation we have $\exp(i{\bf k}.{\bf R})|\psi> = ( |\psi^*><\psi|)|\psi>$, and the interaction Hamiltonian becomes
\begin{equation}
\hat{H}^{EDCM}=G(|e>|\psi ^*><\psi |<g|\hat{a} - |g>|\psi ><\psi^* |<e|\hat{a}^+)
\end{equation} 
We have included CM in the superscript to denote that now the CM variables have been explicitly taken into account.

Finally, we are in position to evaluate the matrix element in Eq. (\ref{eq:die}):
\begin{eqnarray}
 <0_{EM}|_A<\psi ^*|_A<e|\hat{H}_A^{EDCM}|\psi>_A|g>_A|1_{EM}> = \nonumber \\  _A <\psi^*|\psi^* >_A<e|e>_A G_A <0_{EM}|\hat{a}|1_{EM}> =G_0 {\bf e}. {\bf D}_{eg}^A 
\label{eq:onc}
\end{eqnarray}
where the subscript in $G_A$ and the superscript in ${\bf D}_{eg}^A$ have been included to denote that these magnitudes can be different for the two atoms.

With respect to the CM contribution note that for the first term in ${\cal M}_D$ the scalar product of the CM states is one. However, for other terms in ${\cal M}_D$ this contribution is not trivial. For instance, when the initial state is $\varphi$ we have $<\psi^*|\varphi ^*>$. 

We introduce the notation $D_A=D_0  {\bf e}. {\bf D}_{eg}^A $, where $D_0=-it\hbar^{-1}G_0$ contains all the constant coefficients appearing in the above evaluation. Finally, the first term of ${\cal M}_D$ can be expressed as 
\begin{equation}
N_*N_D |a|^2  D_A
\end{equation}
Taking the same steps for the rest of terms in the matrix element we have
\begin{eqnarray}
{\cal M}_D/N_*N_D=D_A+ D_B+ \nonumber \\
2Re(a^*bD_A<\psi^*|\varphi^*><\phi|\chi>) + 2Re(a^*bD_B<\psi|\varphi><\phi^*|\chi^*>)
\end{eqnarray}
The matrix element is a function of the coefficients of the superposition, measuring the entanglement of the system, and the scalar products between the CM states (with and without recoil), determining the non-orthogonality degree. Thus, the absorption rate, $R_D=|{\cal M}_D|^2$, will depend on both, entanglement and non-orthogonality.

For identical particles the calculation was carried in \cite{yap}, but without including the recoil of the absorbing atom. It is immediate to see that when this effect is taken into account the Eq. (10) in that reference becomes:  
\begin{eqnarray}
{\cal M}/2N_fN_I D=2+  2Re(a^*b<\psi^*|\varphi^*><\phi|\chi>)+2Re(a^*b<\psi|\varphi><\phi^*|\chi^*>) \nonumber \\
\pm 2Re(a^*b<\psi^*|\chi^*><\phi|\varphi>) \pm 2Re(ab^*<\varphi^*|\phi^*><\chi|\psi>)  \pm \nonumber \\
2|a|^2Re(<\psi^*|\phi^*><\phi|\psi>)\pm 2|b|^2Re(<\varphi^*|\chi^*><\chi|\varphi>)
\end{eqnarray}
where ${\cal M}$ is the absorption matrix element for identical particles and $D$ is now equal for both atoms. The normalization coefficient for the final state for this case is (Eq. (8) in \cite{yap}):
\begin{eqnarray}
N_f=(4+ 4Re(a^*b<\psi^*|\varphi^*><\phi|\chi>)+4Re(ab^*<\varphi|\psi><\chi^*|\phi^*>) \pm \nonumber \\
4Re(a^*b<\psi^*|\chi^*><\phi|\varphi>) \pm 4Re(ab^*<\varphi^*|\phi^*><\chi|\psi>) \pm \nonumber \\
4|a|^2Re(<\psi^*|\phi^*><\phi|\psi>)\pm 4|b|^2Re(<\varphi^*|\chi^*><\chi|\varphi>))^{-1/2}
\end{eqnarray}

\section{Results}

We graphically represent the above equations to have a clear visualization of the results obtained.  The absorption rates depend on two factors, the coefficients of the two-atom superposition (determining the entanglement degree) and the spatial overlapping of the CM states (responsible for the non-orthogonality in the case of distinguishable atoms and the exchange effects for identical ones). As signalled before we want to represent the absorption rates versus the superposition coefficients. This representation is done for given values of the overlap. In order to specify these values we express all the CM states as a function of a particular basis of the Hilbert space. For the matter of simplicity we assume that the space is two-dimensional, being the extension to higher dimensions straightforward. A basis of the space is implemented by $\psi$ and $\psi ^{\perp}$, with $\psi ^{\perp}$ orthogonal to $\psi$. Other CM states are written as $|\varphi>=c|\psi>+d|\psi^{\perp}>$ and $|\phi>=e|\psi>+f|\psi^{\perp}>$, with $|c|^2+|d|^2=1$ and $|e|^2+|f|^2=1$.  The remaining state can be expressed as $|\chi>=g|\phi>+h|\phi^{\perp}>$ with $|g|^2+|h|^2=1$. Using  $|\phi ^{\perp}>=f|\psi>-e|\psi ^{\perp}>$ we have $|\chi>=(ge+hf)|\psi>+(gf-eh)|\psi^{\perp}>$. For the coefficients $D$ we express them in terms of that of the identical atoms, which consequently are $D=1$. On the other hand, we take $D_A=0.9$ and $D_B=1.1$. We must also define the orthogonal products between the wave functions with recoil. As in general the recoil is small we expect a non very large deviation from the scalar products without recoil. Then we propose an expression of the form $<\eta ^*|\mu ^*>=(0.9 +(1-0.9)<\eta|\mu>)<\eta|\mu>$. The second term in the right hand side of this expression is included to obtain the correct limit when $\mu$  tends towards $\eta$  ($<\eta ^*|\mu ^*> \rightarrow 1$).      

In a first example we consider the two distinguishable atoms having the same degree of non-orthogonality: $<\psi |\varphi>=<\phi|\chi>$, or equivalently, $|\chi>=c|\phi>+d|\phi^{\perp}>$ . All the scalar products can be expressed as a function of $c$, $d$, $e$ and $f$. In figure 1 we represent the case with all the coefficients real, and $c=0.8$ and $e=1/\sqrt 2$.

The main result is the sharp increase of the absorption rate for fermions . It is simple to see that these points correspond to the vicinity of an excluded state. In order to justify this assertion we represent in the right inside graphic of figure 1 the inverse of the squared normalization factor for fermions ($NF=1/N_I^2$ for fermions). We see that the region of sharp increase corresponds to the proximity of the point where the inverse of the normalization factor becomes null. As the normalization factor can be expressed as $N_I^2=1/<\bar{\Phi}|\bar{\Phi}>$, with $\bar{\Phi}$ the unnormalized $\Phi$, we have that $|\bar{\Phi}>=0$ and $|\Phi>=0/0$. This is the typical form of excluded states. 

We could reach the same conclusion by a direct calculation of $\Phi$ for these particular values of the coefficients. It is simple to see that for all values obeying the relation
\begin{equation}
af+bc(gf-eh)-bd(ge+hf)=0
\end{equation} 
we have $|\bar{\Phi}>=0$ and an excluded state. There is a large set of excluded states. We have numerically verified that these states show a sharp increase of the absorption rates similar to that found above.  

In the particular case represented here the absorption increase is of the order of twenty times that of distinguishable and boson atoms. This abrupt increase is an unequivocal footprint of the presence of entanglement-based exclusion. For other values of the coefficient $c$ we observe a similar behaviour. When we have excluded states we see similar peaks, sometimes with larger values and other lower ones, but always in the range of a difference of one order of magnitude. On the other hand, the distinguishable and boson atoms show very similar values  although the form of the curves is different. Note that for non-identical atoms the absorption values are equal for $a=0$ and $a=1$ (absence of multi-particle superposition). In contrast, the boson (and also fermion) values are different in these limiting cases, due to the different overlapping present in both extremes of the interval ($<\psi|\phi> \neq <\varphi |\chi>$).   
\begin{figure}[H]
\center
\includegraphics[width=12cm,height=7cm]{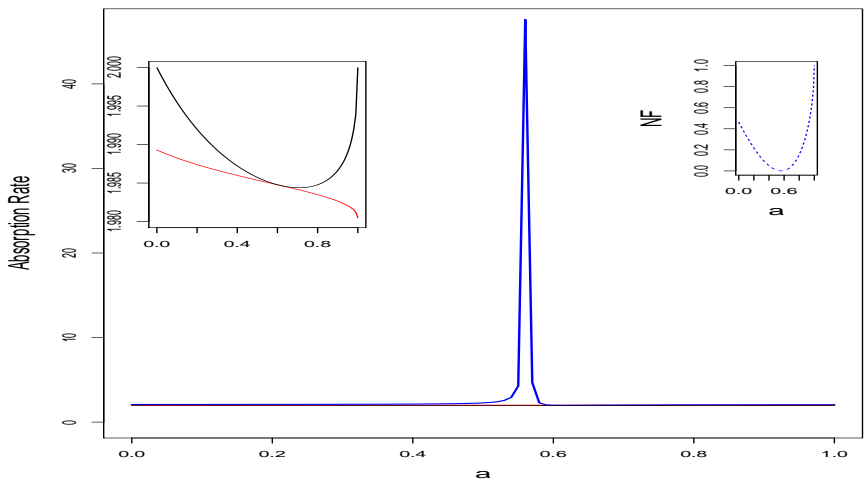}
\caption{Absorption rate in arbitrary units versus the coefficient $a$ of the multi-particle superposition. The black, red and blue lines correspond respectively to distinguishable, boson and fermion pairs of atoms. The left inside graphic represents in a larger scale the two first curves, indistinguishable in the main figure. The right inside graphic sketches the inverse of the squared normalization factor $N_I$ for fermions as a dashed blue curve. }
\end{figure}
It is also interesting to analyse the behaviour of the system when there is not exclusion. We consider the more general case where the non-orthogonality of the two distinguishable atoms can be different. Now, we can have $c \neq g$, and for real coefficients  there are three free parameters, $c$, $e$ and $g$.

For many values of the parameters there is also exclusion and we can observe the peaks described in the previous case, but now we are only interested into the examples without exclusion. In the left side figure we see that all the curves vary with $a$, the coefficient measuring the degree of entanglement or multi-particle superposition. The curves for boson and distinguishable atoms are very similar although with different intensities. The form of the fermion curve is clearly different. Moreover,  the differences with the absorption rates of distinguishable atoms are much larger than for bosons (for $a \approx 0$ of the order of ten times). In the second figure we see a completely different behaviour. Now, for identical atoms there is not dependence on the degree of multi-particle superposition. As in the left side case the absorption rate for fermions is larger than for bosons and distinguishable atoms.
\begin{figure}[H]
\centering
\begin{tabular}{cc}
\includegraphics[width=6cm,height=7cm]{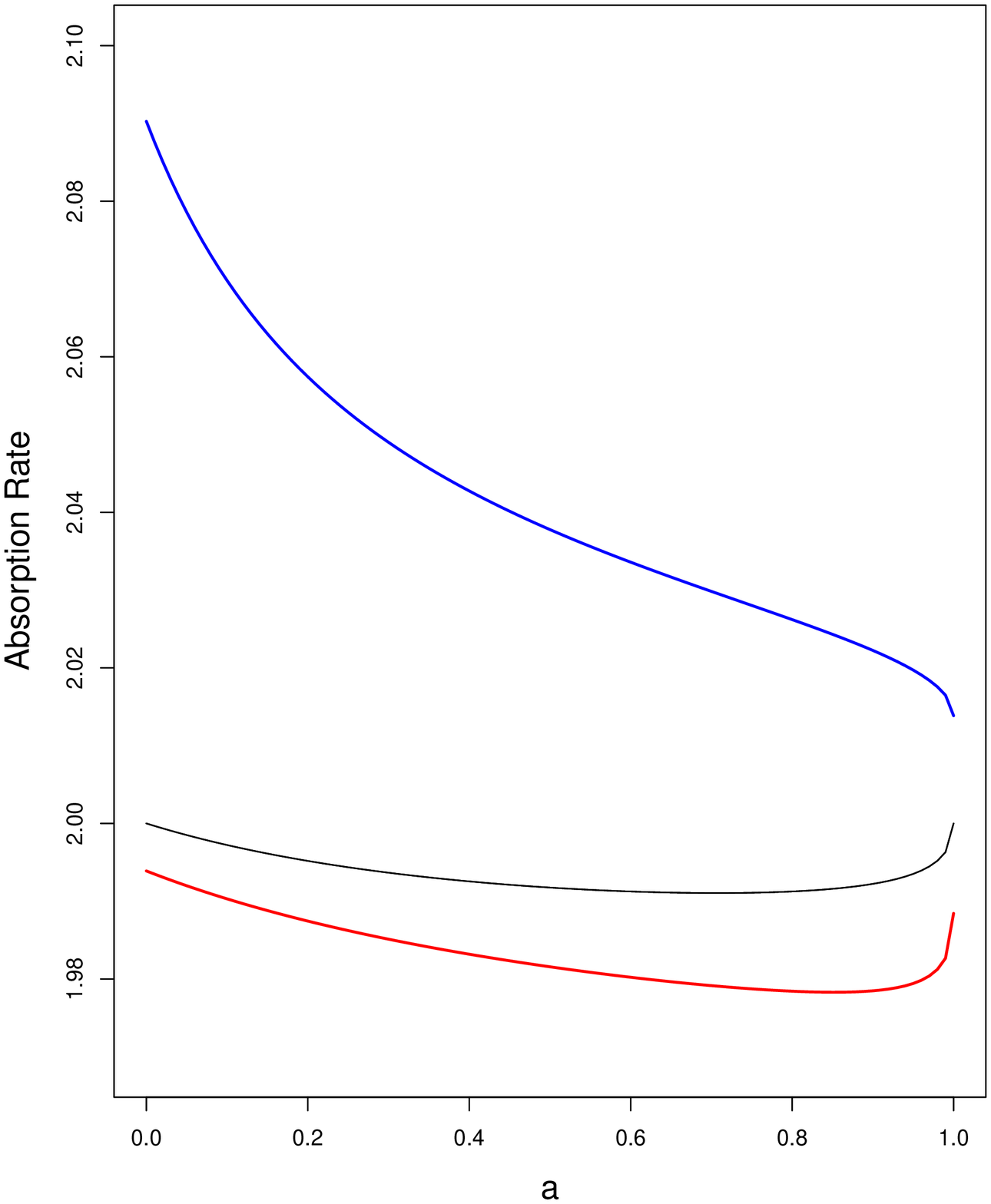}&
\includegraphics[width=6cm,height=7cm]{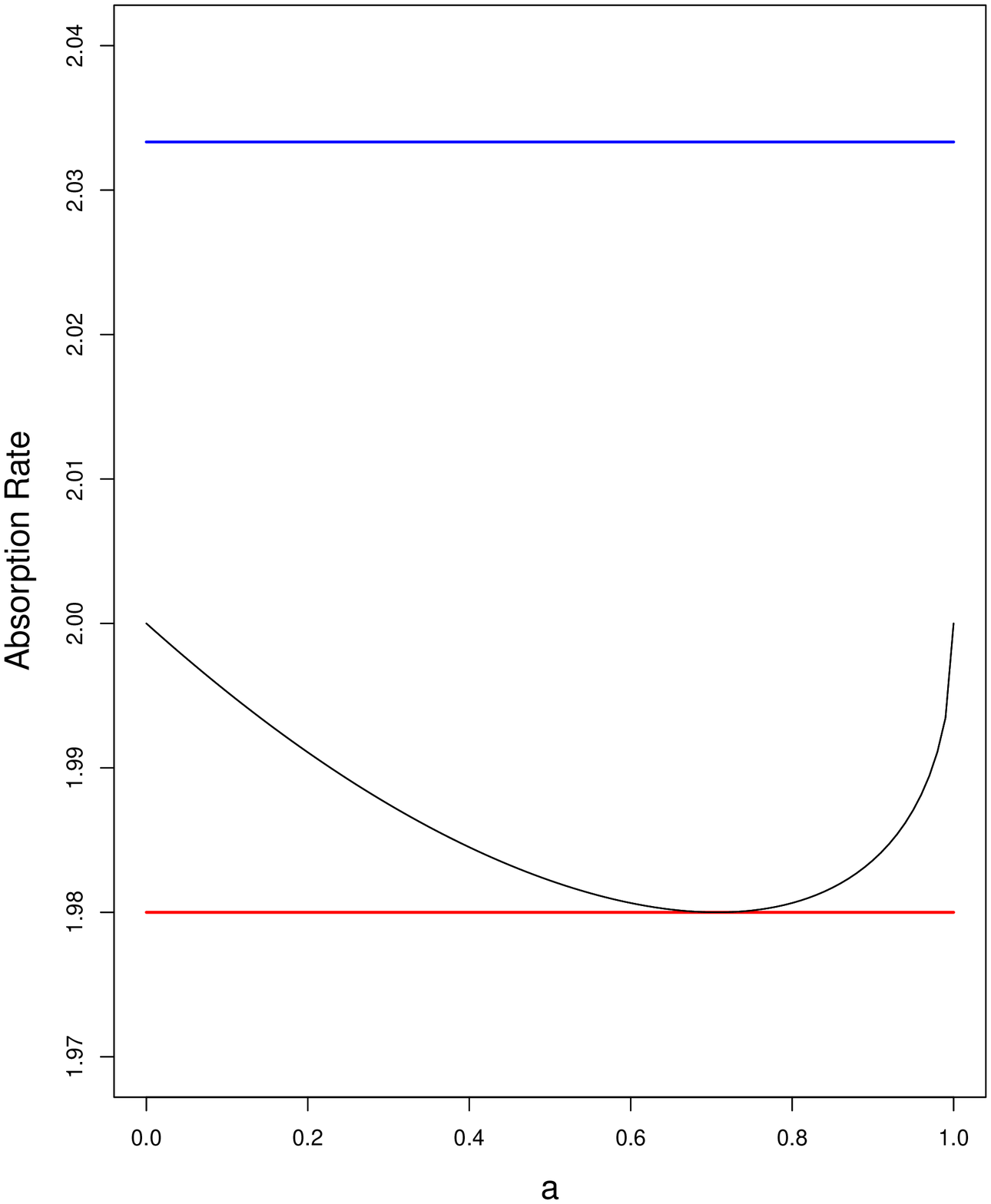}
\end{tabular}
\caption{The same as figure 1 for other values of the CM state coefficients. In the left side we represent the case $c=g=0.9$ and $e=0.3$, and in the right one the parameters are $c=e=g=0.5$.}
\end{figure}

\section{Discussion}

We have revisited the analysis of the joint effects of entanglement and symmetrization when both are simultaneously relevant in the system. The calculations have been improved by the inclusion of the recoil effect, which is important for a correct evaluation of the overlapping and exchange effects after the absorption. In addition, we have compared the results with those of distinguishable atoms instead of mixture states of the same identical atoms.  This allows for a clearer distinction of the contributions of entanglement and of entanglement and symmetrization together. 

We have represented the absorption rates as a function of the coefficients of the two-atom superposition. This way we can directly visualize the dependence of absorption on the entanglement present in the system. We observe a sharp increase in the proximity of the exclusion points. This is an alternative way of illustrating the existence of excluded states. The main advantage of this approach is that in potential experimental implementations of the above ideas the coefficients of the superposition can be manipulated in a simpler way than the overlapping. Moreover, the sharp form of the curve is easier to detect in the presence of noisy data than the flat curve (represented as a function of the overlapping degree) obtained in \cite{yap}, and also seems to be more robust against experimental imperfections.

Another important result emerges from the analysis of the absorption rates. The right hand side graphic of figure 2 shows that, for these values of the parameters, there is no dependence on the coefficients of the two-atom superposition, that is, on the entanglement degree. There is an inhibition of the entanglement effects for identical atoms (they are still present for distinguishable ones). The absorption rates in these cases are only a function of the symmetrization effects (the values of the curves for bosons and fermions are different). The entanglement effects can be cancelled in a multi-particle superposition when symmetrization is simultaneously present. This is a novel manifestation of the rich interplay between entanglement and symmetrization. There is a large variety of behaviours to be explored when both forms of non-separability act at once.

\end{document}